\documentclass[10pt,a4paper,epsf,portrait,times,epsfig]{article}

\usepackage{epsfig}
\usepackage{amsmath}
\usepackage{xspace}
\usepackage{fancybox}
\usepackage{graphicx}
\DeclareGraphicsExtensions{.epsi,.eps,.ps,.eps.gz,.ps.gz,.gif,.epsi.gz}
\def\vckm       {\ensuremath{ {V}_{CKM}}}
\def\FourS {\ensuremath{\Upsilon{(4S)}}\xspace}
\def\lhcb {LHCb\xspace}
\def\invfb   {\ensuremath{\mbox{\,fb}^{-1}}\xspace}
\def\invab   {\ensuremath{\mbox{\,ab}^{-1}}\xspace}
\def\invpb   {\ensuremath{\mbox{\,pb}^{-1}}\xspace}
\def\theckmmatrix  {\ensuremath{ \left( \begin{array}{ccc} V_{ud} & V_{us} & V_{ub} \\ V_{cd} & V_{cs} & V_{cb} \\ V_{td} & V_{ts} & V_{tb} \end{array}\right)}}
\begin{document}
\pagenumbering{arabic}

\begin{center}
{\Large \bf \boldmath{$CP$} Violation in Charm: a New Method}
\end{center}

\vspace{0.5cm} 

\begin{center}
{\large \bf Gianluca Inguglia}\\
\textit{Queen Mary, University of London, Mile End Road, E1 4NS, United Kingdom}\\ 

\textit{g.inguglia@qmul.ac.uk} \\
\vspace{0.5cm}
\today
\end{center}

\vspace{0.5cm}
\begin{center}
{\large \bf Abstract}
\end{center}

We propose for the first time a method to perform analysis of time-dependent $CP$ asymmetries in charm by using both, correlated and un-correlated $D^0$ mesons. Here we consider the decay channels $D^0\to K^+ K^-$  and $D^0\to \pi^+ \pi^-$. The channel $D^0\to K^+ K^-$ will be used to measure the mixing phase, and the difference between the measured phase $D^0\to K^+ K^-$ and $D^0\to \pi^+ \pi^-$ will open the door to the first measurement of $\beta_c$, one of the angle of the charm unitarity triangle. Since in the standard model $CP$ asymmetries in charm are expected to be small, any observation of large time dependent asymmetries or mismatch between predicted and observed value for $ \beta_c$ could signify new physics. We perform and show results of  numerical analysis made considering Super$B$ running at charm threshold, Super$B$ running at $\Upsilon(4S)$ and LHCb and find that Super$B$ and LHCb will be able to measure $\beta_{c,eff}$ with a precision of $1.3^\circ$ and $1.4^\circ$ respectively. The same analysis shows that $\phi_{MIX}$ could be measured at Super$B$ with a precision of $1.3^\circ$.




\section{Introduction}

$CP$ violation represents nowadays one of the most intriguing and challenging fields in physics. Assuming the Big Bang theory as the model that describes the evolution of the early Universe, today we should observe a Universe composed in equal quantities by matter and anti-matter. Since the Universe appears to be dominated in its composition only by matter, a mechanism which has favoured matter over anti-matter had to exist during the first instants of its evolution. Considering the discrete symmetries $C$ (charge symmetry), $P$ (parity), and their combination $CP$, Sakharov in 1967 proposed~\cite{Sak} that the violation of $C$ and $CP$ may have played a very important role in the evolution of the Universe, favoring matter with respect to antimatter. Since the discovery in 1964 of $CP$ violation in the Kaon system~\cite{Fitch}, the research in the field was intensified leading to the discovery of CP violation also in the B system~\cite{CPB}. But in the charm sector this has not been observed, and this represents a missing piece of information in the understanding of flavor dynamics. As it is shown in the literature, for example in~\cite{mcD}, the quantity of $CP$ violation observed in the standard model (SM) of particle physics is not enough to account for the matter anti-matter asymmetry observed in the Universe and new studies are needed in order to look for new sources of $CP$ violation which could explain this asymmetry. This field represents not only a necessary step forward regarding the understanding of elementary particles and flavor dynamics, but it represents a link between particle physics and cosmology.

\section{\boldmath{$CP$} violation in the standard model: the CKM Matrix and the Unitarity Triangle(s)}
In the SM, $CP$ violation naturally arises from a complex phase appearing in Cabibbo-Kobayashi-Maskawa (CKM) matrix~\cite{Cabibbo}~\cite{Kob-Mask}. The matrix is a unitary $3\times3$ matrix which provides a description of quark mixing in terms of the coupling strengths for up to down quark type transitions, and it may be written as

\begin{eqnarray}
\vckm = \theckmmatrix.\label{eq:ckmmatrix}
\end{eqnarray}
Within this framework the probability to observe a transition between a quark $q$ to a quark $q'$ is proportional to $|V_{qq'}|^2$.

\subsection{Wolfenstein and Buras parametrisations of the CKM Matrix}
Different parametrisations of the CKM matrix are available, one of them is the Wolfenstein parametrisation ~\cite{Wolfenstein} shown in Eq.~(\ref{eq:wolfenstein}) which is an expansion in terms of $\lambda = \sin \theta_c$, $A$, $\rho$, and $\eta$, where $\theta_c$ 
is the Cabibbo angle. We have chosen a variant of the Wolfenstein parametrisation elaborated by Buras et al.~\cite{Buras} expanded up to and including terms ${\cal O}(\lambda^5)$ and shown in Eq.~(\ref{eq:ckmmatrixburasrhobaretabar})

\begin{eqnarray}
\vckm =  \ensuremath\left (
 \begin{array}{ccc }
 1 - \lambda^2 / 2      &   \lambda    & A\lambda^3 (\rho - i\eta) \\
 -\lambda  & 1 - \lambda^2/ 2  & A \lambda^2  \\
  A\lambda^3[1-\rho -i\eta]  & -A \lambda^2  & 1
 \end{array}
\right) + {\cal O}(\lambda^4),
\label{eq:wolfenstein}
\end{eqnarray}

\begin{eqnarray}
\vckm =  \ensuremath\left (
 \begin{array}{ccc }
 1 - \lambda^2 / 2 -\lambda^4/8     &   \lambda    & A\lambda^3 (\bar\rho - i\bar\eta)+A\lambda^5(\bar\rho - i\bar\eta)/2 \\
 -\lambda +A^2\lambda^5[1 - 2(\bar\rho+i\bar\eta)]/2 & 1 - \lambda^2/ 2 - \lambda^4(1+4A^2)/8 & A \lambda^2  \\
  A\lambda^3[1-\bar\rho -i\bar\eta]  & -A \lambda^2 + A\lambda^4[1-2(\bar\rho+i\bar\eta)]/2   & 1 -A^2\lambda^4/2
 \end{array}
\right) + {\cal O}(\lambda^6),
\label{eq:ckmmatrixburasrhobaretabar}
\end{eqnarray}

where

\begin{equation} \ensuremath{
  \overline{\rho}= \rho [1- \lambda^2 /2 +  {\cal O}(\lambda^4)],\label{eq:rhobar}  \end{equation}}
\begin{equation} \ensuremath{
\overline{\eta}= \eta [1- \lambda^2 /2 + {\cal O}(\lambda^4)]. \label{eq:etabar} 
\end{equation}}

The choice to consider the CKM matrix with an expansion up to ${\cal O}(\lambda^5)$, is due to the fact that while the expansions to ${\cal O}(\lambda^3)$, have been sufficient for the $B$ factories era, now it becomes necessary to consider additional terms
as we move into the era of LHCb and the Super Flavor Factories (SFF's), not only from a completeness point of view but mainly by another issue: $CP$ violation in charm has to be very small and at ${\cal O}(\lambda^3)$ no CP violation would be predicted. The values of the parameters used in the CKM matrix are shown in Tab.~(\ref{tbl:wolf}).
\begin{table}[!ht]
\caption{Constraints on the Wolfenstein parameters $A$, $\lambda$, $\rho$, $\eta$, $\overline{\rho}$, and $\overline{\eta}$ 
obtained by the UTFit and CKM fitter groups.}\label{tbl:wolf}
\begin{center}
\begin{tabular}{lccc}
Parameter         & UTFit & CKM Fitter & Mean Used \\ \hline \hline
$\lambda$         & $0.22545 \pm 0.00065$ & $0.22543 \pm 0.00077$
                  & $0.22544 \pm 0.00705$
\\
$A$               & $0.8095 \pm 0.0095$   & $0.812^{+0.013}_{-0.027}$
                  & $0.811  \pm 0.015$
\\
$\rho$            & $0.135 \pm 0.021$     & $-$
                  & $-$
\\
$\eta$            & $0.367 \pm 0.013$     & $-$ 

                  & $-$
\\
$\overline{\rho}$ & $0.132 \pm 0.020$     & $0.144\pm 0.025$
                  & $0.138 \pm 0.022$
\\
$\overline{\eta}$ & $0.358 \pm 0.012$     & $0.342^{+0.016}_{-0.015}$ 
                  & $0.350 \pm 0.014$
\\ \hline
\end{tabular}
\end{center}
\end{table}

We adopted the convention to write the CKM matrix in terms of $\overline{\rho}$ and $\overline{\eta}$ because unitarity of the CKM matrix gives rise to six unitarity triangles in the complex plane and $\overline{\rho}$ and $\overline{\eta}$ represent the coordinates of the apex of the $bd$ triangle given in Eq.~(\ref{eq:unitaritytriangle}). Since unitarity triangles are mathematically exact, it is very important to measure their angles and sides to verify unitarity. This allows one to check if the CKM mechanism is the ultimate answer to the $CP$ violation problem in the quark sector or if some new physics (NP) effects have still to be considered. In this sense, a recent article~\cite{Lunghi} has pointed out that the measured value for the $\sin2\beta$ angle of the beauty-unitarity triangle differs by 3.2 standard deviation from the predicted value. In the paper the authors claim that this discrepancy could be interpreted as the fact that CKM mechanism is breaking down.
Two of the six unitarity relationships of the CKM matrix may be written as

\begin{eqnarray}
V_{ud}^* V_{ub} + V_{cd}^* V_{cb} + V_{td}^* V_{tb} = 0, \label{eq:unitaritytriangle}  
\end{eqnarray}
\begin{eqnarray}
V_{ud}^* V_{cd} + V_{us}^* V_{cs} + V_{ub}^* V_{cb} = 0, \label{eq:charmtriangle}  
\end{eqnarray}

where Eq.~(\ref{eq:unitaritytriangle}) is a well known and studied case referred to as the \textit{bd} triangle or unitarity triangle (UT), Eq.~(\ref{eq:charmtriangle}) represents the \textit{cu} triangle that we will call the \textit{charm} unitarity triangle or simply \textit{charm} triangle.

\subsection{The \textit{charm} triangle}
In the previous section we have seen that Eq.~(\ref{eq:charmtriangle}) defines the \textit{charm} unitarity triangle. The angles of the charm unitarity triangle given in Eq.~(\ref{eq:charmtriangle}) are
\begin{eqnarray}
\alpha_c &=& \arg\left[- V_{ub}^*V_{cb}/ V_{us}^*V_{cs} \right], \label{eq:alphac}\\
\beta_c  &=& \arg\left[- V_{ud}^*V_{cd} / V_{us}^*V_{cs} \right], \label{eq:betac}\\
\gamma_c &=& \arg\left[- V_{ub}^*V_{cb} / V_{ud}^*V_{cd}\right]. \label{eq:gammac}
\end{eqnarray}
Using the averages of CKM Fitter and UTFit values for $A$, $\lambda$, $\bar\rho$ and $\bar\eta$ and their errors, we predict that, to order $\lambda^6$
\begin{eqnarray}
  \alpha_c &=& (111.5\pm 4.2)^{\circ}, \\
  \beta_c &=& (0.0350\pm 0.0001)^{\circ}, \\
  \gamma_c &=& (68.4\pm 0.1)^{\circ}.
\end{eqnarray}

\begin{figure}[!ht]
\begin{center}
\resizebox{12.cm}{!}{
\includegraphics{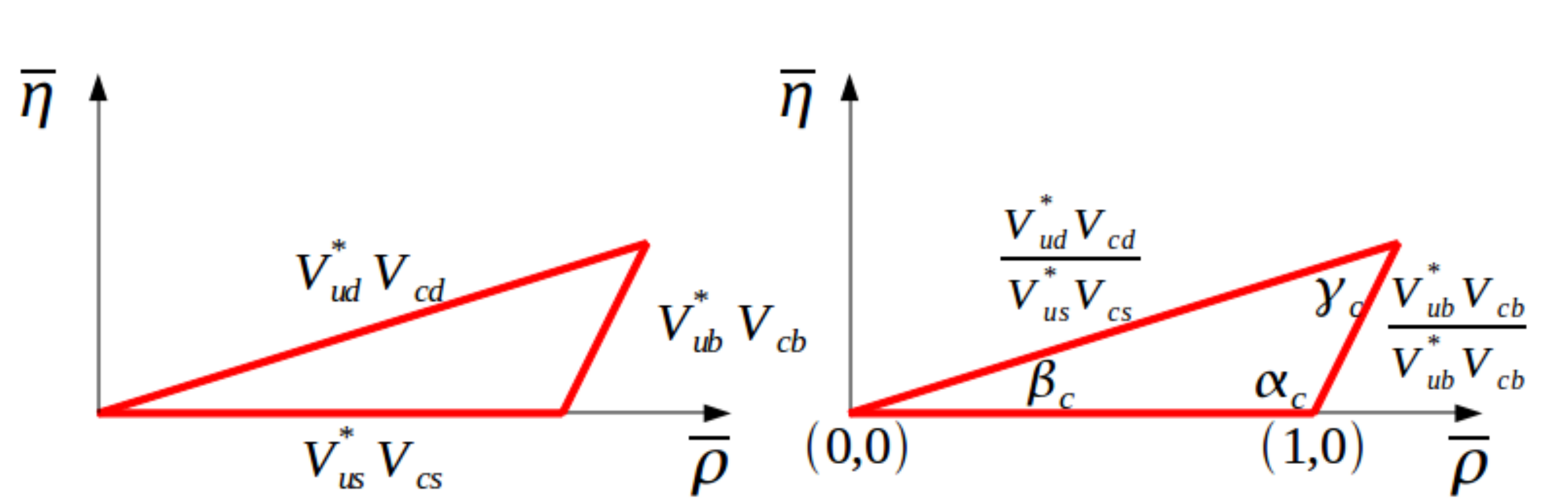}
}
\caption{The \textit{charm} triangle before (left) and after (right) baseline normalization.}\label{fig:charmtriangle}
\end{center}
\end{figure}

These predictions for the angles of the charm triangle could, and should, be 
tested experimentally, either directly (through time-dependent $CP$ asymmetries) 
or indirectly (through measurements of the sides of the triangle).
On comparing Eq.~(\ref{eq:betac}) with Eq.~(\ref{eq:ckmmatrixburasrhobaretabar}), one can 
see that $V_{cd} = V_{cd} e^{i(\beta_c-\pi)}$. The \textit{charm} triangle is shown in Fig.~(\ref{fig:charmtriangle}), where the sides are not in scale.
If one takes the left side of Fig.~(\ref{fig:charmtriangle}) and divides through by $V_{us}^* V_{cs}$, one obtain the triangle shown on the right.
\section{Time-dependent formalism}
Neutral meson systems ($K, D,B_{d,s}$) do exhibit mixing which may be expressed by 
\begin{eqnarray}
i\frac{\partial }{\partial t}
\left( \begin{array}{c} |P^0\rangle \\ |\overline{P}^0\rangle \end{array} \right)
= \textit{$H_{eff}$}
\left( \begin{array}{c}|P^0\rangle \\ |\overline{P}^0\rangle \end{array} \right),\label{eq:tdep:heffmatrix}
\end{eqnarray}
where $|P^0\rangle$ and  $|\overline{P}^0\rangle$ are the strong eigenstates of the mesons and the matrix $H_{eff}= M - i\Gamma /2$ appearing on the right-hand side represents the effective Hamiltonian which describes the mixing. The mass eigenstates may be written as
\begin{eqnarray}
|P_{1,2}\rangle = p|P^0\rangle \pm q|\overline{P}^0\rangle,
\label{eq:tdep:admixture}
\end{eqnarray}
where $q^2+p^2=1$ to normalize the wave function, and 
\begin{eqnarray}
\frac{q}{p} = \sqrt{ \frac{M_{12}^* - i\Gamma_{12}^* / 2} {M_{12} - i\Gamma_{12} / 2} },
\end{eqnarray}
it is important to mention that if $|q/p|\neq 1$ then $CP$ violation in mixing occurs.
\subsection{Time-evolution}
Before discussing the time-dependent formalism it is useful to distinguish between two different cases of $D^0$ meson production: un-correlated and correlated $D^0$ production. Un-correlated $D^0$'s are produced from the decays of $B$ mesons in electron-positron colliders when particles are collided at a center of mass energy corresponding to the $\Upsilon(4S)$ resonance, or from $c\overline{c}$ continuum, or in hadron machines they can be produced both promptly or as decay products of heavier particles. The correlated $D^0$ mesons are instead produced in an electron-positron machine running at a center of mass energy corresponding to the $\Psi(3770)$ resonance. We have obtained the time evolution for both situations described above and it can be shown that these are given by~\cite{bevan-inguglia-meadows}

\textit{(un-correlated case)}
\begin{eqnarray}
\Gamma(P^0 \to f) \propto e^{-\Gamma_1 t}\left[ \frac{\left(1+e^{\Delta\Gamma t} \right)}{2} + \frac{Re(\lambda_{f})}{1+|\lambda_{f}|^2}\left(1-e^{\Delta\Gamma t} \right) + e^{\Delta\Gamma t / 2}\left(\frac{1-|\lambda_{f}|^2}{1+|\lambda_{f}|^2}\cos \Delta M t - \frac{2 Im(\lambda_{f})}{1+|\lambda_{f}|^2}\sin \Delta M t  \right) \right],  \label{EQ:p0toffinal}\\
\Gamma(\overline{P}^0 \to f) \propto e^{-\Gamma_1 t}\left[ \frac{\left(1+e^{\Delta\Gamma t} \right)}{2} + \frac{ Re(\lambda_{f})}{1+|\lambda_{f}|^2}\left(1-e^{\Delta\Gamma t} \right) + e^{\Delta\Gamma t / 2}\left(-\frac{1-|\lambda_{f}|^2}{1+|\lambda_{f}|^2}\cos \Delta M t + \frac{2 Im(\lambda_{f})}{1+|\lambda_{f}|^2}\sin \Delta M t  \right) \right],\label{EQ:p0bartoffinal}
\end{eqnarray}

\textit{(correlated case)}
\begin{eqnarray}
\Gamma(P^0 \to f) &\propto& e^{-\Gamma_1 |\Delta t|}\left[ \frac{h_+}{2} + \frac{ Re(\lambda_{f})}{1+|\lambda_{f}|^2}h_- + e^{\Delta\Gamma |\Delta t| / 2}\left(\frac{1-|\lambda_{f}|^2}{1+|\lambda_{f}|^2}\cos \Delta M \Delta t - \frac{2 Im(\lambda_{f})}{1+|\lambda_{f}|^2}\sin \Delta M \Delta t  \right) \right], \label{EQ:p0toffinaldeltat}\\
\Gamma(\overline{P}^0 \to f) &\propto& e^{-\Gamma_1 |\Delta t|}\left[ \frac{h_+}{2} + \frac{ Re(\lambda_{f})}{1+|\lambda_{f}|^2}h_- + e^{\Delta\Gamma |\Delta t| / 2}\left(-\frac{1-|\lambda_{f}|^2}{1+|\lambda_{f}|^2}\cos \Delta M \Delta t + \frac{2 Im(\lambda_{f})}{1+|\lambda_{f}|^2}\sin \Delta M \Delta t  \right) \right],\label{EQ:p0bartoffinaldeltat}
\end{eqnarray}
where
\begin{eqnarray}
h_{\pm} =  1 \pm e^{\Delta\Gamma |\Delta t|}.
\end{eqnarray}
\\and 
\begin{eqnarray}
\lambda_{f} = \frac{q}{p} \frac{\overline{A}}{A},\label{eq:lambda}
\end{eqnarray}
here $q$ and $p$ are the parameters defining the mixing and $A$ ($\overline{A}$) is the amplitude for the $P$ ($\overline{P}$) decay to a final state $f$. If $|A|^2 \neq |\overline{A}|^2$ there is direct $CP$ violation (in the decay). The study of $\lambda_{f}$ (which should not be confused with the term $\lambda$ appearing in the CKM matrix) is able to probe the combination of $CP$ violation due to mixing and due to decay, and this form of $CP$ violation is referred to as $CP$ violation in the interference between mixing and decay.

\subsection{Time-dependent asymmetries}
Considering Eq.~(\ref{EQ:p0toffinal},\ref{EQ:p0bartoffinal},\ref{EQ:p0toffinaldeltat},\ref{EQ:p0bartoffinaldeltat}) we calculated the time-dependent asymmetries associated to the time evolution of the $D^0$ mesons. The asymmetry is written as follows
\begin{eqnarray}
 {\cal A}(t) = \frac{ \overline{\Gamma} (t) - \Gamma (t) } {\overline{\Gamma}(t) + \Gamma(t) },\label{eq:asym}
\end{eqnarray}

we then obtain \\
\textit{(un-correlated case)}
\begin{eqnarray}
{\cal A}(t) =2 e^{\Delta \Gamma t/2} \frac{ (|\lambda_{f}|^2 - 1)\cos \Delta M t + 2Im \lambda_{f} \sin\Delta M t}{(1 + |\lambda_{f}|^2)(1+e^{\Delta \Gamma t}) + 2 Re\lambda_{f} ( 1 - e^{\Delta\Gamma t})}, \label{eq:asym_unc}\\
\end{eqnarray}
in the limit of $\Delta\Gamma=0$ Eq.~(\ref{eq:asym_unc}) becomes
\begin{eqnarray}
 {\cal A}(t) = -C \cos \Delta M t + S \sin\Delta M t,
\end{eqnarray}
where
\begin{eqnarray}
 S = \frac{2Im \lambda_{f}}{1 + |\lambda_{f}|^2}, \text{ and } C = \frac{1 - |\lambda_{f}|^2 }{1 + |\lambda_{f}|^2}.\label{eq:bzcpparameters}
\end{eqnarray}

\textit{(correlated case)} 

In the case of correlated mesons the asymmetry is obtained, as the time evolution, in terms of $\Delta t$

\begin{eqnarray}
 {\cal A}(\Delta t) = \frac{ \overline{\Gamma} (\Delta t) - \Gamma (\Delta t) } {\overline{\Gamma}(\Delta t) + \Gamma(\Delta t) } =
 2 e^{\Delta \Gamma |\Delta t|/2} \frac{ (|\lambda_{f}|^2 - 1)\cos \Delta M \Delta t + 2Im \lambda_{f} \sin\Delta M \Delta t}{(1 + |\lambda_{f}|^2) h_+ + 2 h_- Re\lambda_{f}}.
 \label{eq:asymdeltat}
\end{eqnarray}

\subsection{Flavor tagging and feasibility of the study in different experimental environments}\label{sec:flavtag}
 In this section we define the different ways to tag the $D^0$ mesons and identify consequently their flavor. The aim of flavor tagging is to identify one (or more) flavor specific final states that can be unambiguously used to determine the flavor of a neutral meson decaying to a CP eigenstate that we want to study. This is a probabilistic assignment and it has consequently an associated probability to be incorrect, which is called mis-tag probability, denoted by $\omega$. This is an important quantity to determine how good a channel is  for flavor tagging. A related quantity to determine the power of a channel for flavor tagging is the dilution $D=1-2\omega$: the closer the dilution is to 1, the better the channel is for flavor tagging. Once again we need to distinguish the two cases of un-correlated and correlated production. 

\textit{(un-correlated case)} 
To flavor tag un-correlated $D^0$ mesons it is necessary to identify a low momentum pion (slow pion) since the process is driven by $D^{*+} \to D^0 \pi^{+}$ and $D^{*-} \to \overline{D}^0 \pi^{-}$. In this way if the the neutral $D$ meson is produced in association with a $\pi^+$ the flavor of the neutral $D$ meson is $D^0$, while if the neutral $D$ meson is produced in association with a $\pi^-$ the flavor of the neutral $D$ meson is $\overline{D}^0$.

\begin{figure}[!ht]
\begin{center}
\resizebox{13.cm}{!}{
\includegraphics{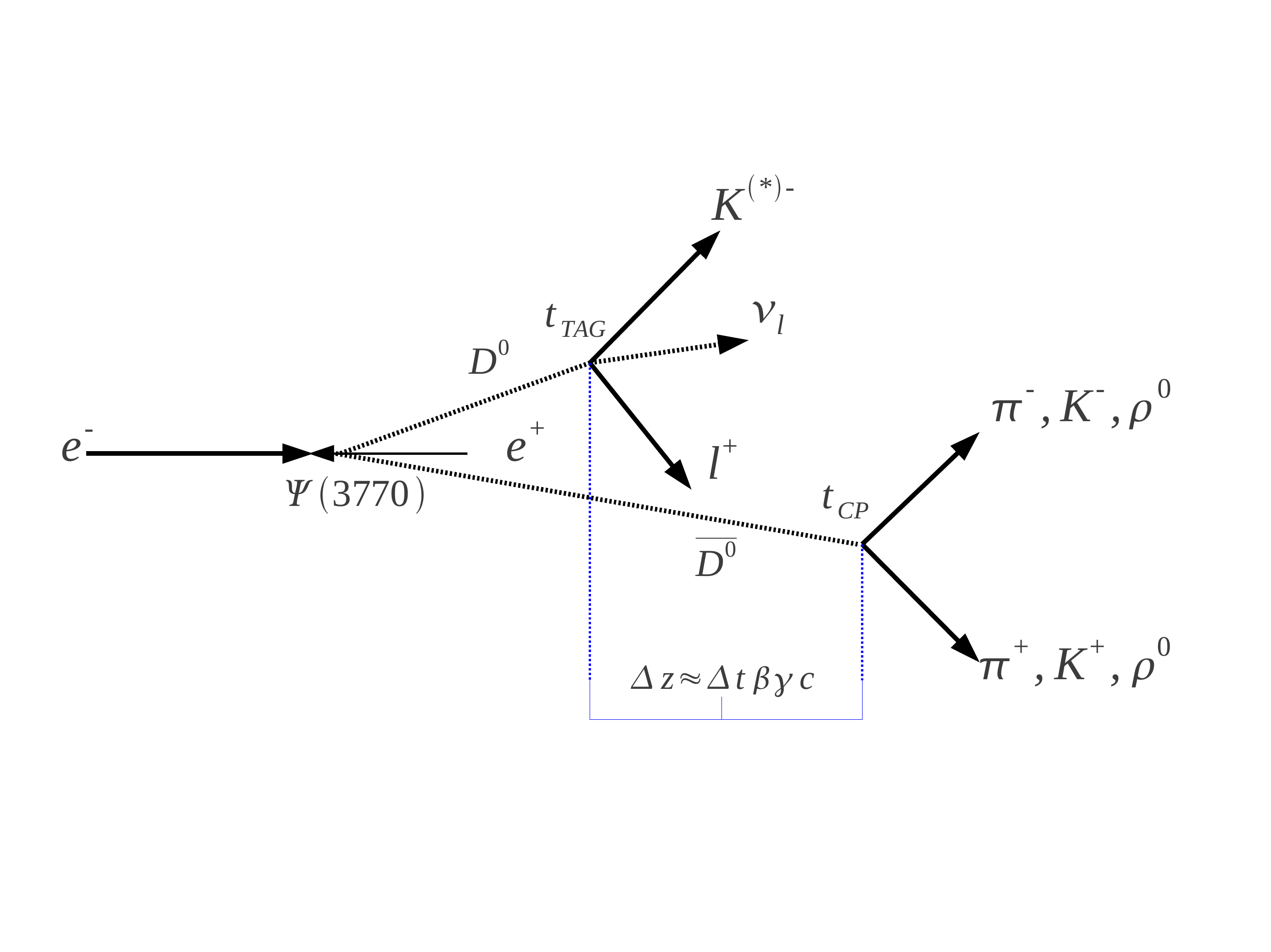}
}
\caption{An example of a semi-leptonic tagged event at \textit{charm} threshold.}\label{fig:tagging}
\end{center}
\end{figure}

\textit{(correlated case)} 
The flavor tag of $D^0$ mesons produced in electron-positron collider running at $\Psi(3770)$ is similar to that to flavor tag $B^0$ mesons produced at $\Upsilon(4S)$. It is possible for example to consider only semi-leptonic decay of the neutral $D^0$ mesons to tag and understand the flavor (Fig.~(\ref{fig:tagging})). The processes $D \to K^{(*)-}l^+\nu$ accounts for 11\% of the D decays. The $D^0$ is associated with a $l^+$ in the final state, the $\overline{D}^0$ is associated to a $l^-$: the flavor is unambiguously assigned, with a mistag probability $\omega$ which is essentially zero.

To evaluate the feasibility of this study we consider now three possible scenarios: a) LHCb, b) Super$B$ (or Belle II) running at the \FourS or c) Super$B$ at the $\Psi(3770)$. We consider the proven performance of $BaBar$~\cite{babar} and Belle and the current performance of LHCb~\cite{lhcb} to estimate the performance in the scenarios that we are going to examine. 
For decays to $CP$ eigenstates (i.e. $D^0 \to K^+ K^-$, $D^0 \to \pi^+ \pi^-$) at LHCb the yield from a sample of 5 \invfb will probably be comparable to that of Super$B$ (Belle II). However since at LHCb hadrons are collided, the levels of background will be higher if compared to those at Super$B$. If one considers the run at charm threshold, then the luminosity will be lower by a factor 10, and the background levels will be lower if compared to those of a run at \FourS. To perform a time-dependent analysis the $D^0$ mesons must to be produced in flight in the laboratory reference frame, and it is then necessary that the flight length of the $D^0$'s exceeds the detector resolution, this will allow one to identify the decay vertex of both $D^0$ $\overline{D}^0$ and perform the analysis. At LHCb $D^0$ mesons are produced both promptly and in $B$ decays, with a momentum of hundreds of $GeV/c$, and it will be important to separate the two different cases. At Super$B$ (Belle II) prompt $D^0$'s from $e^+e^-$ continuum may be easily separated from those from $B$ decays by applying a momentum cut. At charm threshold $D^0$ mesons are produced in a coherent pair from the decay of the $\Psi(3770)$, and they will have a small momentum, consequently the possibility of the measurement of $\Delta t$ strongly relies upon the boost. It is important to mention that Super$B$ will be the first asymmetric electron-positron collider running at charm threshold. Certainly a finer granularity in the time-dependent analysis is possible at LHCb than in the other scenarios. Simulations are needed in order to estimate precision achievable in the three scenarios described above.

\subsection{Mistag probability and dilution: effects on the asymmetry}

We have already mentioned that flavor tagging of neutral mesons is a probabilistic approach to identify the flavor and is affected by mis-tag probability. To account for the mis-tag probabilities one has to consider the physical decay rates instead of the theoretical ones

\begin{eqnarray}
\Gamma^{Phys}(t) &=& (1-\omega)\Gamma (t) + \overline{\omega}\, \overline{\Gamma}(t),\\
\overline{\Gamma}^{Phys}(t) &=& \omega\Gamma (t) + (1 - \overline{\omega})\overline{\Gamma}(t),
\end{eqnarray}
where $\Gamma(t)$ and $\overline{\Gamma}(t)$ are from Eqns.~(\ref{EQ:p0toffinal}) and (\ref{EQ:p0bartoffinal}) and $\overline\omega$ represents the mistag probability for anti-particles. Hence for un-correlated mesons we now can write the $CP$ asymmetry

\begin{eqnarray}
{\cal A}^{Phys}(t) &=& \frac{\overline{\Gamma}^{Phys}(t) - \Gamma^{Phys}(t) } { \overline{\Gamma}^{Phys}(t) + \Gamma^{Phys}(t)}=\Delta \omega + \frac{ (D - \Delta\omega)e^{\Delta \Gamma t/2}[ (|\lambda_{f}|^2 - 1)\cos\Delta M t + 2 Im\lambda_{f} \sin\Delta M t ]}{h_+ (1+|\lambda_{f}|)^2/2 + Re(\lambda_{f}) h_-},
  \label{eq:asymtagging}
\end{eqnarray}

where $\Delta\omega =\omega - \overline\omega$.

With a similar procedure we obtain for the correlated mesons
\begin{eqnarray}
\Gamma^{Phys}(\Delta t) &=& (1-\overline{\omega})\Gamma (\Delta t) + \omega \, \overline{\Gamma}(\Delta t),\label{eq:taggedrates}\\
\overline{\Gamma}^{Phys}(\Delta t) &=& \overline{\omega}\, \Gamma (\Delta t) + (1 - \omega)\overline{\Gamma}(\Delta t),
\label{eq:bartaggedrates}
\end{eqnarray}
where $\Gamma(\Delta t)$ and $\overline{\Gamma}(\Delta t)$ are now from Eqns.~(\ref{EQ:p0toffinaldeltat}) and (\ref{EQ:p0bartoffinaldeltat}), and

\begin{eqnarray}
{\cal A}^{Phys}(\Delta t) = \frac{\overline{\Gamma}^{Phys}(\Delta t) - \Gamma^{Phys}(\Delta t) } { \overline{\Gamma}^{Phys}(\Delta t) + \Gamma^{Phys}(\Delta t)}= -\Delta \omega + \frac{ (D + \Delta\omega)e^{\Delta \Gamma |\Delta t|/2}[ (|\lambda_{f}|^2 - 1)\cos\Delta M\Delta t + 2 Im\lambda_{f} \sin\Delta M \Delta t ]}{h_+ (1+|\lambda_{f}|)^2/2 + Re(\lambda_{f}) h_-}.
  \label{eq:asymtagging1}
\end{eqnarray}

\section{CP eigenstates}

In our study we have considered a sample of 35 two and three body $CP$ eigenstate decay channels of $D^0$ mesons and we  determined the CKM matrix elements contributions to the decay amplitudes~\cite{bevan-inguglia-meadows}. From these we evaluated the information on the weak phase that may be extracted by each decay. Here we concentrated on two of these channels: $D^0 \to K^+ K^-$ and $D^0 \to \pi^+ \pi^-$.
We are interested in the value of $\lambda_{f}$ and when exploring $CP$ violation, this may be written as
\begin{eqnarray}
\lambda_{f} = \left| \frac{q}{p}\right|e^{i\phi_{MIX}} \left| \frac{\overline{A}}{A}\right|e^{i\phi_{CP}},
\end{eqnarray}
where $\phi_{MIX}$ is the phase of the mixing between $D^0$ and $\overline{D}^0$ mesons, $\phi_{CP}$ the overall $CP$ phase related to decay of the $D^0$ meson to specific final $CP$ eigenstate $f_{CP}$, $\overline{A}$ and $A$ represent the amplitude of the decay for $\overline{D}^0$ and $D^0$. The amplitude $A$ ($\overline{A}$) is determined by the contribution of different decay topologies and consequently it is not always related to one angle of the \textit{charm} triangle. To write the amplitude one has to consider that tree (T), color suppressed tree (CS), weak exchange (W), and penguin ($P_{q}$) topologies (amplitudes) may contribute to the decay channel
\begin{eqnarray}
A &=& |T|e^{i(\phi_T+\delta_{T})} + |CS|e^{i(\phi_{CS}+\delta_{CS})} + |W|e^{i(\phi_{W}+\delta_{W})} + \sum\limits_{q=d, s, b} |P_q|e^{i(\phi_{P_{q}}+\delta_{P_{q}})},\label{eq:amplitude}
\end{eqnarray}
where (for $j=T,CS,W,P_q$) $\phi_{j}$ are the weak phases (which change sign under $CP$ transformations) of the different topologies, the moduli in front to the exponentials are their magnitude and $\delta_{j}$ are the strong phases (which are invariant under $CP$ transformations) associated to each amplitude. 
\begin{table*}[!ht]
\caption{$CP$ eigenstate modes considered in this paper indicating the topologies
contributing to each process in terms of the CKM factors associated with 
$T$ (tree), $P_{q}$ (penguin where $q$ is a down-type quark),
and $W_{EX}$ (W-exchange) transitions.  
Blank entries in the table denote that a given topology does not 
contribute to the total amplitude of the decay, and the relative strengths of these amplitudes
decrease from left to right (see ~\cite{bevan-inguglia-meadows} for the full table).}
\label{tbl:eigenstates}
\begin{center}
\begin{tabular}{lcccc}
mode & $\eta_{CP}$ & $T$ &  $P_{q}$ & $W_{EX}$ \\ \hline \hline
$D^0\to K^+ K^-$             & $+1$ & $V_{cs}V_{us}^*$ &   $V_{cq}V_{uq}^*$  & \\
$D^0\to \pi^+ \pi^-$         & $+1$ & $V_{cd}V_{ud}^*$ &   $V_{cq}V_{uq}^*$  & $V_{cd}V_{ud}^*$ \\
\hline
\end{tabular}
\end{center}
\end{table*}

The decays $D^0 \to K^+ K^-$ and $D^0 \to \pi^+ \pi^-$ are tree dominated and assuming negligible other contributions (this would be adequate for a first rudimentary time-dependent CP asymmetry measurement), it follows that

\begin{eqnarray}
\lambda_{f} = \left| \frac{q}{p}\right|e^{i\phi_{MIX}} e^{-2i\phi_{T}^W},
\end{eqnarray}

where $|T|$ and the strong phase $e^{i\delta_T}$ cancel in the ratio of $\overline{A}/A$.
It is clear that if one wants to relate the CP asymmetry to one of the angles of the \textit{charm} triangle, then the role of the penguin contribution must to be understood and a precision measurement of $q/p$ must be obtained.
If now we consider the elements of Table~\ref{tbl:eigenstates} we can evaluate the products of the CKM factors involved in the different processes, and these are:

\begin{eqnarray}
 V_{cs}V_{us}^* &=& \lambda - \frac{\lambda^3}{2} - \left(\frac{1}{8} +\frac{A^2}{2}\right) \lambda^5,\\
 V_{cd}V_{ud}^* &=& -\lambda + \frac{\lambda^3}{2} + \frac{\lambda^5}{8} + \frac{A^2\lambda^5}{2}[1 -2(\bar\rho +i\bar\eta)],\\
 V_{cb}V_{ub}^* &=& A^2 \lambda^5(\bar\rho + i\bar\eta),
\end{eqnarray} 

Two of these three amplitudes are complex, $V_{cb}V_{ub}^*$ and $V_{cd}V_{ud}^*$. $V_{cb}V_{ub}^*$ has a large phase ($\gamma_{c}$), while $V_{cd}V_{ud}^*$ is related to a small phase ($\beta_{c}$).

\subsection{$D^0 \to K^+K^-$}
\label{sec:dzerotokk}

The channel $D^0\to K^+K^-$ measures the phase of $V_{cd}V_{ud}^*$ only in a sub-dominant penguin transition (of the order of $\lambda^5$), and is otherwise dominated by a real tree amplitude with a CKM factor of $V_{cs}V_{us}^*$ (of the order of $\lambda$).  
Hence to first order one would expect to observe an asymmetry consistent with the mixing
phase $\phi_{MIX}$, with no CKM weak phase contribution.  This channel provides 
measurements of $|q/p|$ and $\phi_{MIX}$ to complement others that may be available.
Given that the SM prediction of the asymmetry in this channel is small, this
is also an ideal mode to use when searching for NP.
It is interesting to note that $V_{cs}$ is complex at ${\cal O}(\lambda^6)$
 using the convention of ~\cite{Buras}.  A measurement 
of $\beta_c$ could be possible, this is not the most promising mode to measure 
the angle.

\subsection{$D^0 \to \pi^+\pi^-$}
\label{sec:dzerotopipi}

$D^0\to \pi^+\pi^-$ measures the phase of $V_{cd}V_{ud}^*$ in the leading order tree, one of the penguin
amplitudes, and the $W$ exchange topologies.  Of the remaining two penguin amplitudes
that contribute to this decay, one is completely negligible (mediated by a $b$ quark
loop) and the other is of the order of $\lambda$.  The non-trivial penguin 
topologies are doubly Cabibbo suppressed loops and proceed at order $\lambda^2$,
where as the tree amplitude is singly Cabibbo suppressed.  A rudimentary measurement
of this process could in principle ignore the penguin contribution, in which case
$Im\lambda_{f} \simeq \sin(\phi_{MIX} - 2\beta_c)$.  Thus there will be a four-fold 
ambiguity in any measurement of $\beta_c$. However one should note that a 
more complete analysis would be required in order to extract the weak
phase and disentangle the contribution from the $c\to s \to u$ penguin.

\section{Time-dependent \boldmath{$CP$} violation: numerical analysis}
In this section the three possible experimental scenarios described in Section~\ref{sec:flavtag} are compared.
For the numerical analysis and the extrapolation to the expected precision on $\beta_{c}$ 
we have generated a set of hundred Monte Carlo data samples assuming $SuperB$ running 
at charm threshold (semi-leptonic tagging plus kaon tagging), Super$B$ running at 
\FourS and the LHCb case, respectively. We generated events according to the time 
dependences described in Sec. III, and we perform a binned fit to the resulting 
asymmetry using Eq.~(\ref{eq:asymtagging}) and Eq.~(\ref{eq:asymtagging1}), where $\arg(\lambda_f)$ and $|\lambda_{f}|^2$
are allowed to vary in the fit to data. 
This analysis was repeated for different possible values of the phase (i.e. $\arg(\lambda_{f})=0 ^\circ, 10 ^\circ $, etc.). 
The fit results are then expressed in terms of the two parameters and their relative errors and are used to compute the achievable sensitivity on $\beta_{c}$, as described in Sec.~\ref{sec:sensitivity}.

\textit{(Super$B$ at $\Psi(3770)$)} 
$D$ meson pairs produced at the $\psi(3770)$ are quantum-correlated, so that the time evolution 
is given by Eqns.~(\ref{EQ:p0toffinaldeltat}) and (\ref{EQ:p0bartoffinaldeltat}).  If one accounts
for tagging dilution, then the time-dependent $CP$ asymmetry is given by Eq.~(\ref{eq:asymtagging1}).  
On restricting time-dependent analysis to using only semi-leptonic tagged decays, the asymmetry 
will simplify as there is no dilution, hence both $\omega$ and $\Delta \omega$ terms can be 
neglected, and any systematic uncertainty on the asymmetry arising from $D\simeq 1$ will be small.  
Furthermore the $e^+e^-\to \psi(3770)$ environment
is extremely clean, so that any systematic uncertainty from background contributions will be small
and under control.  These are important points to stress as we know that the $CP$ phase of interest 
is small, hence in order to make a precision measurement the systematic uncertainties must be minimized.

With 500\invfb of data at charm threshold one can expect to accumulate approximately $1.89 \times 10^9$
$D$ meson pairs.  We anticipate that Super$B$ could record $158,000$ $X^+ e \nu_e$ tagged events, 
corresponding to 489500 events when using the full set of $K^{(*)} \ell \nu_\ell$ tagged events, $\ell = e, \mu$~\cite{bevan-inguglia-meadows}.
In order to estimate the precision with which one can measure 
$\arg(\lambda_{f})$, we perform unbinned likelihood fits to a number of simulated experiments, each
consisting of $D^0$ and $\overline{D}^0$ tagged signal events, with distributions in $\Delta t$
given by Eqns.~(\ref{eq:taggedrates}) and (\ref{eq:bartaggedrates})  and asymmetry given by Eq.~(\ref{eq:asymtagging1}) with $\omega=\Delta\omega=0$.
The average uncertainty $\arg(\lambda_{f})$ is taken to represent the typical
result of a given experiment. 
This numerical estimate only includes semi-leptonic tagged events.  One can also consider
using hadronically tagged events, for example $D^0\to K^- X$ ($K^+ X$), 
where $X$ is anything, which correspond to 54\% (3\%) of all $D^0$ meson decays.  From
these modes alone, one would expect $\omega \simeq 0.03$, and that the asymmetry in
particle identification of $K^+$ and $K^-$ in the detector will naturally lead to 
a small, but non-zero value of $\Delta \omega$.  We expect that there would be 
approximately 2.2 million kaon tagged $D^0\to \pi^+\pi^-$ events in 500\invfb 
at charm threshold.  Using these data alone, one would be able to measure
$\phi$ to a precision of $4^\circ$, hence if one combines the results from 
semi-leptonic and kaon tagged events, a precision of $\sigma_\phi\sim 3.4^\circ$
is achievable.

\textit{(Super$B$ at \FourS)} 
The scenario at the \FourS is somewhat more complicated than
the situation encountered at the $\psi(3770)$.  Firstly in order to remove background from
$D$ mesons produced in $B$ meson decay, one restricts the analysis to mesons with high momentum.
In addition to non-trivial backgrounds, one also has to consider non-zero tagging dilution, where
the asymmetry is that given in Eq.~(\ref{eq:asymtagging}).
Thus it is not obvious that $\Delta \omega$ can be neglected, and indeed $D \neq 1$.  
We estimate that one could reconstruct $6.6 \times 10^6$ tagged $D^0\to \pi^+\pi^-$ events
in a data sample of 75\invab with a purity of 98\%.   

\textit{(LHCb)} 
The final scenario considered is that of measuring time-dependent asymmetries from uncorrelated
$D$ mesons in a hadronic environment, where dilution and background effects 
will be larger than at an $e^+e^-$ machine.  Given that the measurement of $\lambda_{f}$ is 
expected to be dominated by systematic uncertainties, it is not clear what the ultimate precision obtained 
from LHCb will be.  The best way to ascertain this would be to perform the measurement.  
We estimate that LHCb will collect 
$7.8 \times 10^6$ $D^*$ tagged $D^0 \to \pi^+\pi^-$ decays in 5\invfb of data, based on an initial
37\invpb of data.  Based on the data shown in the reference, we estimate a purity of $\simeq 90\%$ and
$\omega \simeq 6\%$.

\subsection{Sensitivity on $\beta_{c}$} \label{sec:sensitivity}
We report here the summary of the sensitivity estimates made on the basis of our simulations.
More accurately, since $D^0\to K^+K^-$ measures the mixing phase and $D^0\to \pi^+\pi^-$ measures $\phi_{MIX} + 2\beta_{c,eff}$ the difference between phases of $\lambda_{f}$ measured in $D^0\to K^+K^-$ and $D^0\to \pi^+\pi^-$
decays is $\phi_{CP} =  -2\beta_{c,eff}$.
If loop contributions can be well measured and both long-distance and weak 
exchange contributions are negligible, then this constraint can be translated into 
a measurement of $\beta_c$.  The corresponding sensitivity estimates for the 
different scenarios considered are summarized in Table~\ref{tbl:numericalanalysis}.
We estimate that it should be possible to measure $\phi_{CP}$ to 
$\sim 2.6^\circ$ using this approach~\cite{bevan-inguglia-meadows}. Assuming that penguin contributions can be measured precisely,
then the error on $\beta_{c,eff}$ that could be obtained by Super$B$ would be $\sim 1.3^\circ$. LHCb will
require input from Super$B$ on the decay modes with neutral particles in the final state in order 
to translate a measurement of $\beta_{c,eff}$ to one on $\beta_c$.  
Further work is required to understand how penguins and other suppressed amplitudes 
affect the translation of $\beta_{c,eff}$ to $\beta_c$,
however it is clear that there will be a significant contribution from penguins given the
size of the $D^0\to \pi^0\pi^0$ branching fraction.

We determined the phase $\arg(\lambda_{f})$, but it is worth mentioning that we are also able to constrain $|\lambda_{f}|$ with the same set of measurements.  If one observes
$|\lambda_{f}| \neq 1$ in data it would constitute a measurement of direct $CP$ violation in a given decay 
channel.  We estimate that it should be possible to measure $|\lambda_{f}|$ with a statistical uncertainty 
of $1-4\%$ at future experiments such as Super$B$. Something interesting to note is
that Super$B$ proponents expect to accumulate 500\invfb of data at charm threshold in only
three months, whereas 75\invab would require five years of running at nominal luminosity. We also obtain the expected precision for the mixing phase (Table~\ref{tbl:numericalanalysisMIX}) at Super$B$ to be $\pm 1.3^\circ$ when running at center of mass energy corresponding to the \FourS and $\pm 2.1^\circ$ when running at the $\Psi(3770)$ center of mass energy and using semi-leptonic and kaon tagged events.

\begin{table}[!ht]
\caption{Summary of expected uncertainties from 500\invfb of data at charm threshold, 75\invab of 
data at the \FourS, and 5\invfb of data from \lhcb for $D^0\to\pi^+\pi^-$ decays.
The column marked SL corresponds to semi-leptonic tagged events, and 
the column SL+K corresponds to semi-leptonic and kaon tagged events at charm threshold.}
\label{tbl:numericalanalysis}

\begin{center}
    \begin{tabular}{ | l | l  l  l |l|}
    \hline
    Parameter & SL & SL + K & $\Upsilon(4S)$ & LHCb\\ \hline \hline
    $\phi = \arg(\lambda_{f})$ & $8.0^\circ$  & $3.4^\circ$ & $2.2^\circ$ & $2.3^\circ$  \\  
    $\phi_{CP}=\phi_{KK}-\phi_{\pi\pi}$ & $9.4^\circ$ & $3.9^\circ$ & $2.6^\circ$ & $2.7^\circ$  \\ 
    $\beta_{c,eff}$ & $4.7^\circ$ & $2.0^\circ$ & $1.3^\circ$ & $1.4^\circ$\\
    \hline
    \end{tabular}
\end{center}
\end{table}

\begin{table}
\caption{Summary of expected uncertainties from 500\invfb of data at charm threshold and 75\invab of 
data at the \FourS for $D^0\to K^+K^-$ decays.}\label{tbl:numericalanalysisMIX}
\begin{center}
    \begin{tabular}{ | l | l  l  l |}
    \hline
    Parameter & SL & SL + K & $\Upsilon(4S)$ \\ \hline \hline
    $\phi_{MIX} (\phi_{KK})$ & $4.8^\circ$  &   $2.1^\circ$ & $1.3^\circ$   \\  
        \hline
    \end{tabular}
\end{center}
\end{table}

The method proposed here represents a new way to measure the mixing phase, since the current method to measure the mixing phase is via a time-dependent Dalitz Plot analysis of D decays to self conjugate final states.

\section{Conclusions}
We have explored and defined for the first time the formalism required to perform time-dependent $CP$ asymmetries in charm decays
using correlated $D^0 \overline{D}^0$ decays as well a $D^0$ mesons tagged from $D^*$
decays, and we discussed two interesting channels: $D^0 \to \pi^+\pi^-$ and $D^0 \to K^+K^-$. One can use 
$K^+K^-$ decays to measure the mixing phase precisely and other decays can be used 
to constrain the angle $\beta_{c,eff}$ which is one of the angles of the \textit{charm} triangle.  
A data sample of 500\invfb collected at charm threshold could be collected at Super$B$ in only three months and it would provide a 
sufficient test to constrain any potential large NP effects.
Similar measurements would be possible using $D^*$ tagged decays at Super$B$, Belle II and
LHCb.  We expect the statistical precision on the measured 
phase at Super$B$ to be slightly better than the one achievable at  LHCb. 
The measurements
proposed here provide a new set of consistency checks of CKM that can be performed using $D$ decays.  
Measurements of the sides of the triangles would
enable us to perform an indirect cross-check of CKM.  
Super$B$ has a potential advantage over other experiments 
as it will be able to collect data at charm threshold with a boosted center 
of mass, as well as being able to explore effects using neutral mesons from 
a $D^*$ tagged event.  Data from charm threshold will be almost pure, with 
a mis-tag probability of $\sim 0$ for semi-leptonic tagged events, which 
could be advantageous if systematic uncertainties dominate measurements
from \FourS data and from LHCb.  Within our formalism a measurement of $| \lambda_{f} | \neq 1$ could also signify direct $CP$ violation. 

\section{Acknowledgments}

The work presented here is a contribution to the proceedings of the \textit{"$\textit{49}^{ th}$ Course of the International School of Subnuclear Physics"} held in Erice (Italy). This study has been supported by Queen Mary, University of London and it is based on a previous study done in collaboration with Dr. A.J. Bevan (Queen Mary, University of London) and Prof. B. Meadows (University of Cincinnati) who I wish to thank for their teaching and suggestions. I wish to thank the Ettore Majorana Foundation and Centre for Scientific Culture and Prof. A. Zichichi for inviting me at the event and for giving me the possibility to share this work with the community.

\end{document}